# Integrated multi-color Raman microlasers with ultra-low pump levels in single high-Q lithium niobate microdisks

Guanghui Zhao[1,5], Jintian Lin[1,6,†] Botao Fu[1,5], Renhong Gao[1], Chuntao Li[2,3], Ni Yao[4], Jianglin Guan[2,3], Minghui Li[1,6], Min Wang[2], Lingling Qiao[1], and Ya Cheng[1,2,3,5,*]

[1] *State Key Laboratory of High Field Laser Physics and CAS Center for Excellence in Ultra-Intense Laser Science, Shanghai Institute of Optics and Fine Mechanics (SIOM), Chinese Academy of Sciences (CAS), Shanghai 201800, China*

[2] *The Extreme Optoelectromechanics Laboratory (XXL), School of Physics and Electronic Science, East China Normal University, Shanghai 200241, China*

[3] *State Key Laboratory of Precision Spectroscopy, East China Normal University, Shanghai 200062, China*

[4] *Research Center for Humanoid Sensing, Zhejiang Lab, Hangzhou 311100, China*

[5] *School of Physical Science and Technology, ShanghaiTech University, Shanghai 200031, China*

[6] *Center of Materials Science and Optoelectronics Engineering, University of Chinese Academy of Sciences, Beijing 100049, China*

†*jintianlin@siom.ac.cn*

*\*ya.cheng@siom.ac.cn*

Photonic integrated Raman microlasers, particularly discrete multi-color lasers which are crucial for extending the emission wavelength range of chip-scale laser sources to much shorter wavelength, are highly in demand for various spectroscopy, microscopy analysis, and biological detection. However, integrated multi-color Raman microlasers have yet to be demonstrated because of the requirement of high-Q microresonators possessing large $\chi^{(2)}$ nonlinearity and strong Raman phonon branches and the challenging in cavity-enhanced multi-photon hyper-Raman scattering parametric process. In this work, integrated multi-color Raman lasers have been demonstrated for the first time at weak pump levels, via the excitation of high-Q ($>6\times10^6$) phase-matched modes in single thin-film lithium niobate (TFLN) microresonators by dispersion engineering. Raman lasing was observed at 1712 nm for a 1546-nm pump threshold power of only 620 µW. Furthermore, multi-color Raman lasers were realized at discrete wavelengths of 1712 nm, 813 nm, 533 nm and 406 nm with pump levels as low as 1.60 mW, which is more than two order of magnitude lower than the current records (i.e., 200 mW) in bulk resonators, allowed by the fulfillment of the requisite conditions consisting of broadband natural phase match, multiple-resonance and high Q-factors.

## 1. Introduction

Photonic integrated Raman lasers based on stimulated Raman scattering (SRS), have been considered as one of the effective means to extend the available spectral coverage of conventional lasers, offering new laser signals at any desired wavelength within the transparent window of Raman-active media with high compactness, low energy consumption, and scalability.[1-5] To achieve integrated Raman lasers with low pump levels in the continuous-wave (CW) regime, whispering gallery microresonators are favored to construct integrated Raman lasers due to the high-Q factors and small volume for dramatically reducing the pump powers and enabling photonic integration.[1-5] Recent developments of microresonator-based integrated Raman microlasers have been demonstrated in various photonic integrated platforms, such as silicon,[6-8] diamond,[9] aluminum nitride,[2] lithium niobate,[3,4,10,11] silicon carbide,[12] silica,[1,5,13] and chalcogenide,[14] through the leveraging of the non-parametric process by aligning the pump wavelength and Stokes Raman vibrational frequency to the cavity modes (i.e., double resonance). To further extend the available spectral coverage to much shorter wavelength under single-wavelength CW pumping, it is necessary to exploit the stimulated multi-photon hyper-Raman scattering (SMPHRS)[15,16] for realizing integrated multi-color lasers. Unlike SRS based on $\chi^{(3)}$ non-parametric process, SMPHRS is dominated by cascaded parametric process between the pump wave and the generated SRS signal through $\chi^{(2)}$ sum-frequency generation. Therefore, achieving SMPHRS requires two prerequisites to be met simultaneously in terms of material platforms and cavity-enhanced optical parametric process. First, the material platforms should possess strong Raman-active phonon branches, large $\chi^{(2)}$ nonlinearity, broad transparent window, as well as the potential for high-performance photonic integration. Luckily, thin-film lithium niobate (TFLN) [17-20] is such a platform, and has been used to fabricate whispering gallery microresonators with Q factor higher than $10^7$.[11,21] And second, the requisite condition for cavity-enhanced parametric process consists of the fulfillment of broadband phase match, and multiple-resonance conditions with large spatial modal overlap to trigger SMPHRS. However, this challenging is difficult to be overcome simultaneously. Therefore, integrated multi-color Raman lasers have not yet been demonstrated in single microresonators.

Here, we demonstrate integrated multi-color Raman microlasers in high-Q TFLN microdisks with low pump levels by dispersion engineering under single-wavelength CW optical pumping at 1547 nm. SMPHRS is achieved by designing and tailoring the dispersion properties to fulfill the broadband phase match and multiple resonance conditions. And the dispersion engineered TFLN microresonators were fabricated with Q factors higher than $6\times10^6$ for compensating the relatively small (> 1 %) spatial modal overlap for promoting the efficient nonlinear interplay. The pump threshold for SRS reaches as low as 0.62 mW, thanks to the high-Q factors and small mode volume. And integrated multi-color Raman laser signals are observed for the first time, with discrete spectral wavelengths of 1712, 812, 533, and 406 nm at pump power of only 1.60 mW, which is more than two order of magnitude lower than the current records (200 mW) in bulk resonators.[15,16]

## 2. SMPHRS in the single TFLN microdisk

### 2.1 Fabrication of the integrated TFLN microdisk

X-cut TFLN wafer which consisting of 0.7 μm thick TFLN, 2 μm thick silica insulator layer, and a lithium niobate handle, was chosen to fabricate the microstructures. The TFLN wafer was fabricated to suspended TFLN microdisks by femtosecond laser photolithography assisted chemo-mechanical etching (PLACE) technique,[11] with a diameter of 54.7 μm to support plenty of spatial whispering gallery mode (WGM) families. Figure 1(b) shows the fabricated TFLN microdisk, indicating an ultra-smooth surface.

### 2.2 The experimental setup

The experimental setup for the formation of multi-color Raman microlasers is schematically plotted in Fig. 1(a). A tunable narrow-linewidth (< 200 kHz) laser (Model TLB-6728, New Focus Inc.) with output power tuned by an inline variable optical attenuator (VOA) was used as pump light source. The pump light was coupled into the microdisk by an optical tapered fiber with 2 μm waist diameter. The position of the tapered fiber can be accurately controlled by a 3D piezo-electric stage with a solution of 20 nm. Here, the distance between the microdisk center and the tapered fiber was chosen as 26.6 μm to excite high-order WGMs in the microdisk. An optical microscope imaging system consisting of an objective lens with numerical number of 0.42, optional optical filters, and infrared (IR)/visible charge coupled device (CCD), was amounted above the microdisk to monitor the coupled position, and capture the excited spatial modal intensity distribution. The polarization of the input light was adjusted as transverse-electrical (TE) polarization with an inline polarization controller. The generated nonlinear signals in the microdisk were coupled out of the microdisk by the same tapered fiber. The output signal from the tapered fiber was sent to a photodetector (PD) and optical spectrometers for transmission spectrum and optical spectrum analysis, respectively. To collect and analyze the discrete multi-color Raman laser signals with spectral coverage ranging from the visible band to the IR band, three kinds of spectrometers, including a spectrometer with detection spectral range from 350 nm to 1000 nm, two optical spectral analyzers (detection spectral range from 600 nm to 1700 nm, detection spectral range from 1200 nm to 2400 nm) were employed.

**2.3 Integrated multi-color Raman lasers at low pump levels**

**2.3.1 Forward single-longitudinal-mode Raman laser**

When the pump wavelength $\lambda_p$ is set as 1547.0 nm with a pump power of more than 0.70 mW, forward stimulated Raman Stokes line $\lambda_R$ located at 1712 nm, corresponding to the Raman shift of 690 cm$^{-1}$,[22] is detected in the optical spectrum, as shown in Fig. 2(a). Figure 2(b) plots the Raman laser peak power varying with the increasing pump power, showing a linear dependency with a slope efficiency of 32.7%. The laser power threshold was measured to 0.62 mW, as indicated in Fig. 2(b), which is defined as the critical input power above which the Raman line peak power linearly grows as the input power increases.[23] The inset of Fig. 2(b) shows the captured spatial mode intensity profile of the IR signals emitted from the microdisk, exhibiting high-order WGM characteristic. The maximum peak power reaches 295 µW when further increasing the pump power to 1.60 mW, showing a side-mode suppression ratio as high as 50.4 dB, as shown in Fig. 2(c). Remarkably, when the pump power raised to more than 1.60 mW, cascaded Raman scattering was detected at 1914.3 nm. Moreover, Raman-assisted four wave mixing signals were generated around 1712.3 nm wavelength with one free spectral range of ~8.30 nm. The spectrum is plotted in Fig. 2(d) at the pump power of 3 mW. The cascaded Raman scattering and Raman-assisted four wave mixing would impede the further increase of the single Raman line, and lead to an unstable Raman laser. Therefore, to endure a relatively stable Raman laser emission, the pump power is limited to be not more than 1.60 mW.

**2.3.2 Multi-color Raman lasers through SMPHRS**

Multi-color Raman laser signals accompanied with harmonics generation were detected at the pump power of 1.60 mW, which is confirmed by the spectrum shown in Fig. 3(a) and the bright spatial mode intensity profile captured by the visible CCD shown in Fig. 3(b). The near IR hyper-Raman spectral line appears at 812.7 nm wavelength. And the output power of this Raman line is measured as 36.30 µW, which is close to the power 36.36 µW of the generated second harmonic generation (SHG) signal. This maximum power of SHG is lower than the measured threshold pump power 620 µW at the pump wavelength. In consideration of the loaded Q factors of the SHG mode is often lower than that of the pump mode,[24] the SHG power is not high enough to directly trigger stimulated Stokes Raman line with Raman frequency shift of 690 cm$^{-1}$ because the threshold pump power for SRS is inversely quadratically proportional to the loaded Q factor.[23] Therefore, this Raman spectral line at 812.7 nm matches well with the stimulated two-photon hyper-Raman line (HR I) of $\frac{1}{1/\lambda_p + 1/\lambda_R}$, other than SRS direct from SHG. The spatial mode intensity profiles of the two-photon hyper-Raman mode and SHG are captured with 800 nm long-pass filters and bandpass filters by visible CCD, which are shown in Figs. 3(c) and (d), respectively. Both the intensity profiles exhibit high-order WGM characteristic. Besides this hyper-Raman line HR I, there are also two visible hyper-Raman spectral lines located at 532.8 nm (HR II) and 406.1 nm (HR III) observed in the spectrum of Fig. 3(a), which match well with three-photon Raman lines of $\frac{1}{2/\lambda_p + 1/\lambda_R}$ and $\frac{1}{2/\lambda_p + 2/\lambda_R}$. And the corresponding output powers of these two emission lines are 6.84 μW and 0.73 μW, respectively. Additionally, cascaded third harmonic generation (THG) at 515.3 nm is also detected with an output power of 25.46 μW. The captured spatial modal intensity profile at wavelength range shorter than 600 nm is plotted in Fig. 3(e), indicating its high-order WGM characteristic as well as all the other intensity profiles. Although these spatial intensity profiles are not identical, the faultiness can be compensated by the ultra-high Q factors to enhance the nonlinear interplay, allowed by the dramatical suppression of the scattering loss by the use of PLCAE technique.[11,21] Thus, on-chip efficient multi-color Raman laser has been demonstrated in the TFLN microdisk at a weak pump level of only 1.60 mW, which is more than 2 order of magnitude lower than the best results (i.e., 200 mW) previously reported in bulk resonators,[15,16] thanks to much smaller mode volume.

**3 Revealing of the mechanics underlying the parametric processes**

**3.1 SHG at low pump power < 0.6 mW**

Since the SMPHRS is the complicated optical parametric processes, it is necessary to study the simple parametric process like SHG to reveal the universal physical mechanics underlying the bright multi-color Raman laser. The pump power was controlled to be less than 0.6 mW to avoid triggering the SRS. Second harmonic signal at wavelength of 773.5 nm was detected, as shown in Fig. 4(a). The conversion efficiency of second harmonic for varied pump powers, is shown in Fig. 4(b), indicating a linear growth and a high conversion efficiency of 22.1%/mW. And the loaded Q factors of the participated modes are characterized. The transmission spectra of the pump mode and the second harmonic mode are shown in Figs. 4(c) and (d), indicating the loaded Q factors are as high as $6.14 \times 10^6$ and $1.26 \times 10^6$, respectively. These ultra-high Q factors are one of the key parameters to significantly promote the nonlinear conversion efficiencies, and result in a dramatical reduction of the threshold pump power for SRS.

## 3.2 Analyzation of the broadband phase match and multiple resonance

Besides high-Q factors, the realization of the integrated multi-color Raman laser relys on broadband phase match and multiple-resonance conditions to trigger SMPHRS and harmonics generation.[24,25] These abundant parametric processes are originated from broadband phase match, which can be expressed as[26]

$$m_1 + m_2 - m_3 = m_3, \qquad (1)$$

where $m_{mc}$ represents the angular momentum carried by the nonlinear polarization of the microdisk, and $m_1$, $m_2$, and $m_3$ are the angular momentums of the pump, SRS, and multi-photon Raman modes, respectively. Here, the angular momentum of the mode, $m = n \cdot cos\Theta \cdot kr$, is determined by the effective refractive index $(n)$ of the TFLN, the angular momentum $(m)$ of the mode, the ratio of the angular momentum of the microdisk to the total momentum $(cos\Theta)$, the wave vector $(k)$, and the radius of the microdisk $(r)$. Because the microdisk hosts a variety of WGMs that exhibit nearly-continuous geometric features, this inherent characteristic of the microdisk enables natural phase match over a wide range of frequency at a cost of partial sacrifice of radial modal overlap factors, as shown in Tab. 1. The spatial modal overlap factors for SHG and THG were calculated to be 2.67% and 1.35%, respectively. It should be noted that the fabricated TFLN microdisk possesses ultra-high Q factors. These ultra-high-Q phase-matched modes are leveraged to realize the efficient SMPHRS for compensating the relatively low modal overlap factors by increasing the efficient interplay length.

**Table 1.** The participating phase-matched modes for broadband phase matched parametric processes.

| $\lambda(\mu m)$ | 1.7123 | 1.5470 | 0.8127 | 0.7735 | 0.5328 | 0.5153 |
|---|---|---|---|---|---|---|
| $n$ | 1.992 | 2.019 | 2.187 | 2.202 | 2.295 | 2.308 |
| $\cos\Theta$ | 0.966 | 0.968 | 0.893 | 0.890 | 0.851 | 0.847 |

## 4. Conclusion

To conclude, we have demonstrated bright on-chip multi-color Raman lasers for the first time. This multi-color microlaser is operated at a pump level as low as 1.60 mW through the excitation of ultra-high-Q phase matched modes in the single TFLN microdisk under single CW telecom laser pump. Such integrated multi-color coherent source is doomed to promote future advancements of integrated white lasers and fast optical information processing.

**Acknowledgements**

The authors thank Prof. Haisu Zhang at East China Normal University for the helpful discussion.

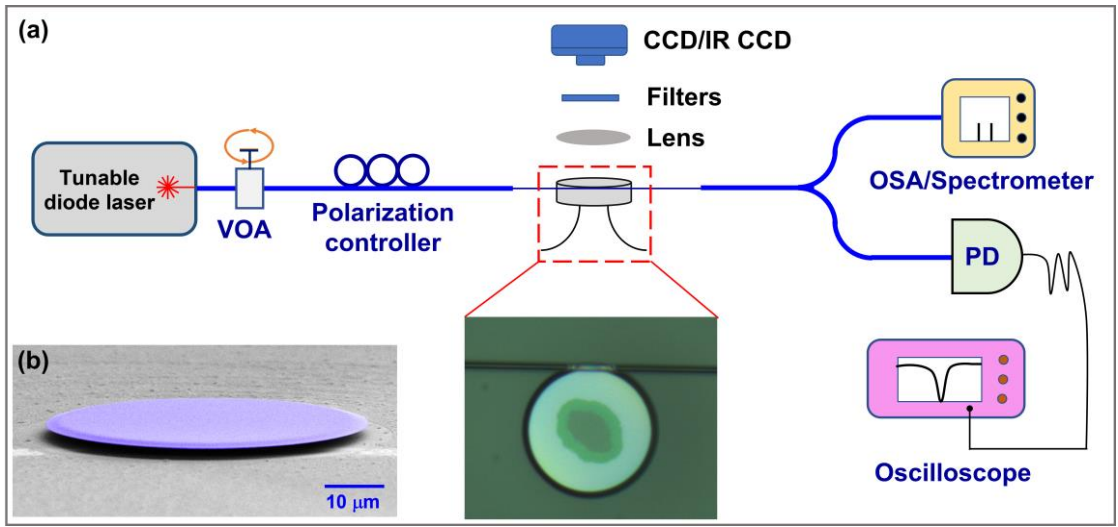

**Figure 1.** Demonstration of the integrated multi-color Raman laser. a) Experimental setup for Raman lasing and mode distribution characterization. Optical spectral analyzer is denoted as OSA. Inset: Optical micrograph of the tapered fiber coupled with the X-cut TFLN microdisk. b) Scanning electron microscope (SEM) image of the microdisk.

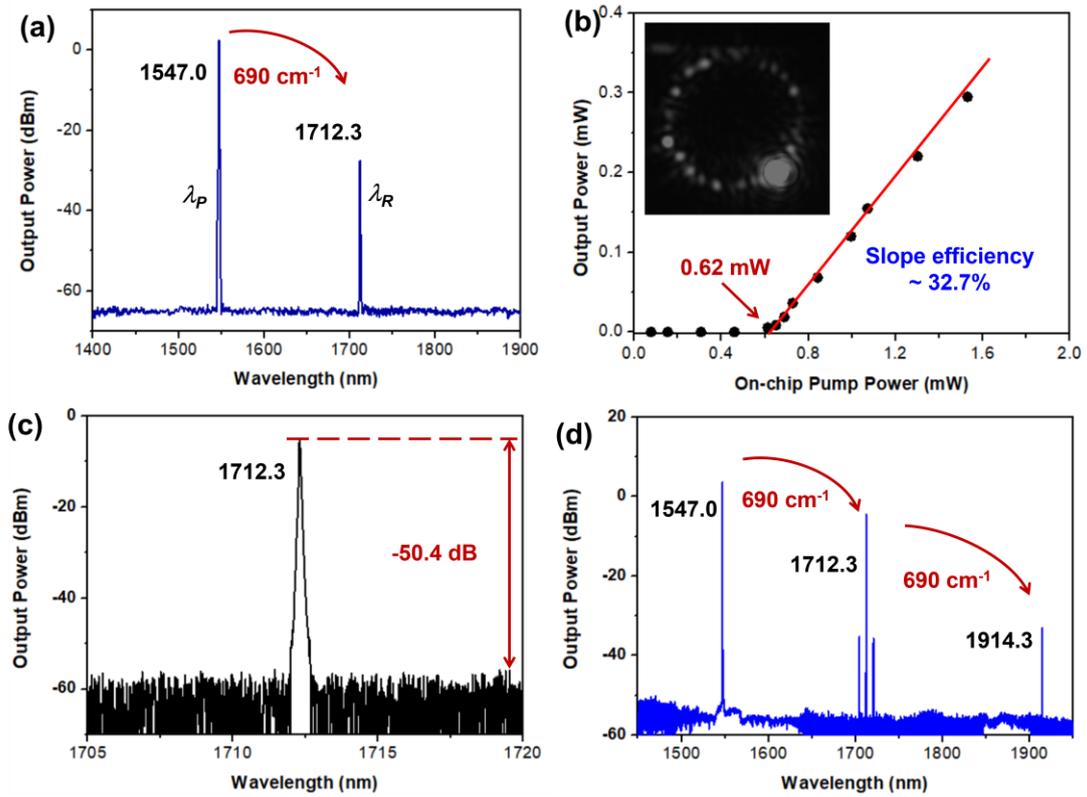

**Figure 2.** The forward single longitudinal mode Raman laser. a) The spectrum of the single-frequency Raman laser. b) The Raman peak power dependence on the pump power. Inset: Optical micrograph of the IR emission from the microdisk. c) The side mode suppression of the Raman laser. d) The spectrum of the cascaded Raman laser signal and Raman assisted four-wave mixing signals at pump power of 3 mW.

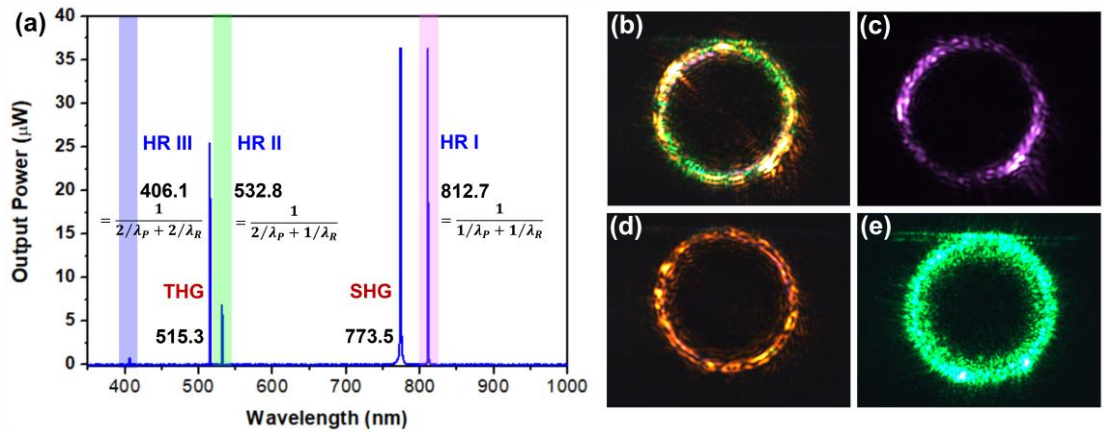

**Figure 3.** Realization of the integrated multi-color Raman laser. a) The spectrum of the multi-color Raman laser accompanied with harmonics generation. Here, multi-photon hyper-Raman lines are denoted as HR I - HR III. Optical micrographs of b) the entire nonlinear signal emission with wavelength range shorter than 1000 nm from the microdisk captured by visible CCD, c) the HR I emission, d) the SHG emission, and e) the nonlinear signal emission with wavelength range shorter than 600 nm.

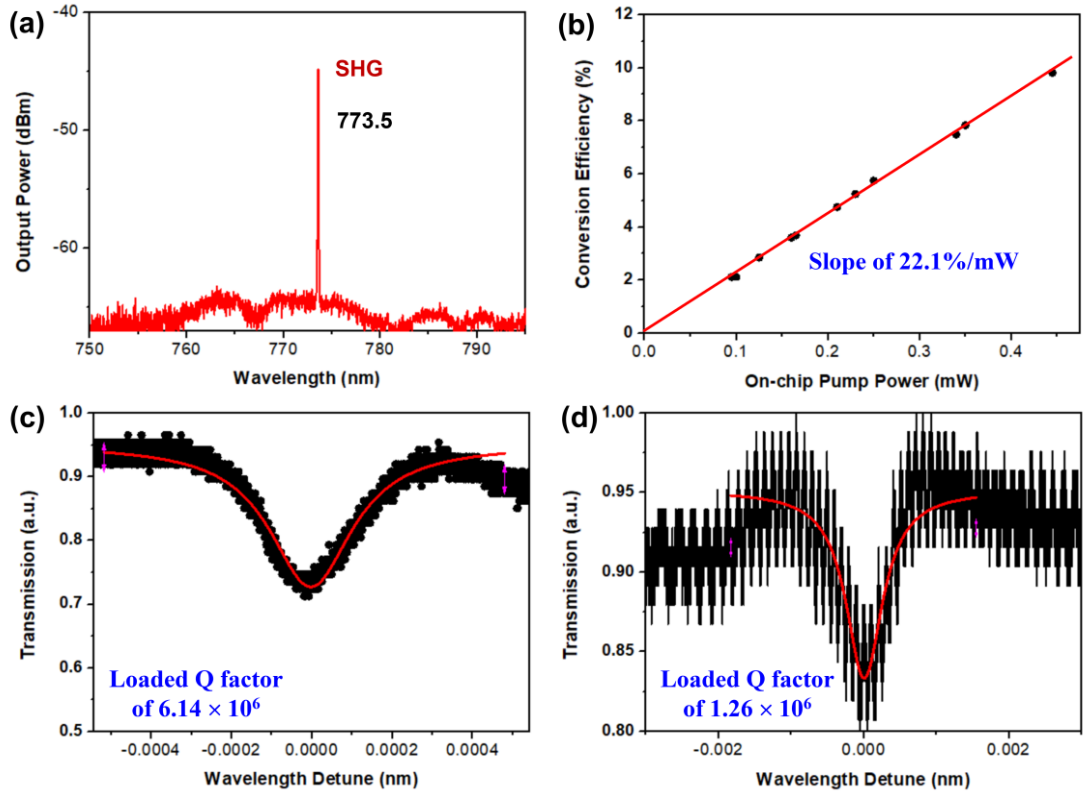

**Figure 4.** Second harmonic generation. a) Spectrum of the second harmonic signal. b) Conversion efficiency of the second harmonic signal as a function of the pump power. c) Loaded Q factor of the pump mode. d) Loaded Q factor of the second harmonic mode.